\documentclass[cits]{PoS}

\PoS{PoS(LAT2005)326}

\title{Eigenmodes of covariant Laplacian in SU(2)
Yang--Mills vacuum: higher representations%
\addtocounter{footnote}{1}\thanks{Our research is supported in part
by the U.S. Department of Energy under Grant No.\ DE-FG03-92ER40711
(J.G.), the Slovak Science and Technology Assistance Agency under
Contract No.\ APVT-51-005704 (\v{S}.O.), grants RFBR 04-02-16079,
RFBR 05-02-16306-a (M.I.P.),  grants RFBR 05-02-16306-a, RFBR
05-02-17642, and an Euler-Stipendium (S.N.S.).}}
\ShortTitle{Eigenmodes of covariant Laplacian in SU(2) Yang--Mills
vacuum: higher representations}
\author{J.\ Greensite\\
        The Niels Bohr Institute, Blegdamsvej 17, DK-2100 Copenhagen \O, Denmark;\\
        Physics and Astronomy Dept., San Francisco State University,
        San Francisco, CA~94117, USA\\
        E-mail: \email{greensit@stars.sfsu.edu}}
\author{\v S.\ Olejn\'\i k\\
        Institute of Physics, Slovak Academy of Sciences,
        SK--845 11 Bratislava, Slovakia\\
        E-mail: \email{stefan.olejnik@savba.sk}}
\author{M.\ I.\ Polikarpov, \addtocounter{footnote}{-2}\speaker{S.\ N.\ Syritsyn}\\
        ITEP, B.\ Cheremushkinskaya 25, Moscow 117259, Russia\\
        E-mail: \email{polykarp@itep.ru}, \email{syritsyn@itep.ru}}
\author{V.\ I.\ Zakharov\\
        Max-Planck Instit\"ut f\"ur Physik, F\"ohringer Ring 6,
        D-80805 Munich, Germany\\
        E-mail: \email{xxz@mppmu.mpg.de}}

\abstract{The study of lowest eigenmodes of the covariant Laplacian
in fundamental representation of the gauge group revealed their
specific localization properties. These may bear information on
confinement of fundamental scalar particles in SU(2) Yang--Mills
vacuum. It was expected that scalar particle eigenmodes in other
representations would be localized in different physical volumes.
However simulations show strikingly different results for the
adjoint and higher ($J=3/2$) representations. Apart from much higher
extent of localization, we find evidence of different scaling
behavior of localized eigenmodes.}

\FullConference{XXIIIrd International Symposium on Lattice Field Theory\\
                 25-30 July 2005\\
                 Trinity College, Dublin, Ireland}

\newcommand{\beq}{\begin{equation}}
\newcommand{\eeq}{\end{equation}}
\newcommand{\bdm}{\begin{displaymath}}
\newcommand{\edm}{\end{displaymath}}
\newcommand{\beqn}{\begin{eqnarray}}
\newcommand{\eeqn}{\end{eqnarray}}
\newcommand{\bea}[1]{\beq\begin{array}{#1}}
\newcommand{\eea}{\end{array}\eeq}
\newcommand{\bi}{\begin{itemize}}
\newcommand{\ei}{\end{itemize}}
\newcommand{\ben}{\begin{enumerate}}
\newcommand{\een}{\end{enumerate}}
\newcommand{\bc}{\begin{center}}
\newcommand{\ec}{\end{center}}

\newcommand{\Tr}{\mathrm{Tr}}

\newcommand{\LQ}{\Lambda_{QCD}}

\newcommand{\hmu}{\hat{\mu}}

\newcommand{\lp}{\left}
\newcommand{\rp}{\right}

\newcommand{\IPR}{\mathrm{IPR}}

\newcommand{\pghalf}{0.50\textwidth}
\newcommand{\schalf}{0.6}
\newcommand{\schalfps}{0.434}

\begin{document}

\section{Introduction}

%In \cite{locpaper,JGtalk} properties of covariant Laplacian eigenmodes were
%studied. Making use of lattice regularization the quantum Yang--Mills field
%with SU(2) gauge group was simulated with Monte-Carlo. Those field
%trajectories were taken as a background for a quenched scalar particle
%introduced by a covariant Laplacian.  The eigenstates for a color particle
%in a fundamental representation at least for the bottom of the spectrum were
%localized, which could be interpreted as a signal of confinement. Other
%cases of a scalar particle with open color, namely, of adjoint and $J=3/2$
%color representattions, also showed localization of lowest eigenmodes but
%the localization is on infinitesimal volume in continuum limit. In this
%report the study is reviewed and extended to the case of finite
%temperatures: above and below of decomfinement transition.

    A preceding talk at this meeting~\cite{JGtalk} presented our
results~\cite{locpaper} on localization properties of the covariant
Laplacian in the fundamental representation. The operator studied is
the simplest discretization of the covariant Laplacian:
\beq \lp(\Delta\phi\rp)^\alpha_x = \sum_{\mu}\lp[
    U_{x,\mu}^{\alpha\beta}\phi^\beta_{x+\hmu}
    - 2 \phi^\alpha_x +
    U_{x,-\mu}^{\alpha\beta}\phi^\beta_{x-\hmu}\rp],
\eeq
where $U^{\alpha\beta}_{x,\mu}$ is covariant transporter in the
given representation.

    To investigate localization we calculate the Inverse Participation
Ratio of the probability density of a wave function:
\beq \IPR = {V \sum_x \rho^2 (x) \over \lp(\sum_x \rho (x) \rp )^
2}\,, \quad \rho(x) = \phi^\dag(x) \phi(x) \eeq
which allows us to estimate the ``mean'' localization volume at given
parameters and reveal its scaling properties.

\section{Adjoint representation}

\begin{figure}[b!]
\begin{minipage}{\pghalf}\bc
$V_{loc} \sim a^{\mathbf 2} $ \\
\includegraphics[scale=\schalf]{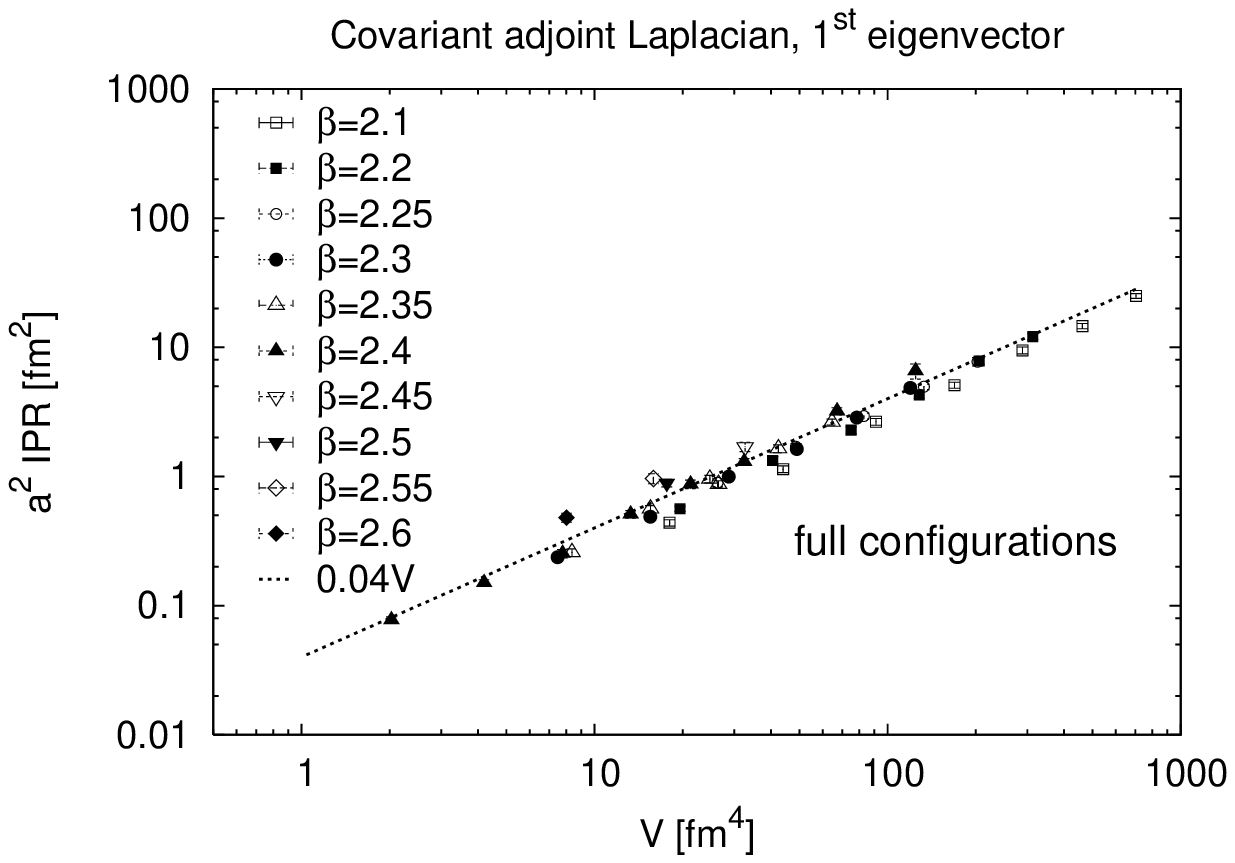}
\ec\end{minipage}
\begin{minipage}{\pghalf}\bc
$V_{loc} / a^2  \approx const $ \\
\includegraphics[scale=\schalf]{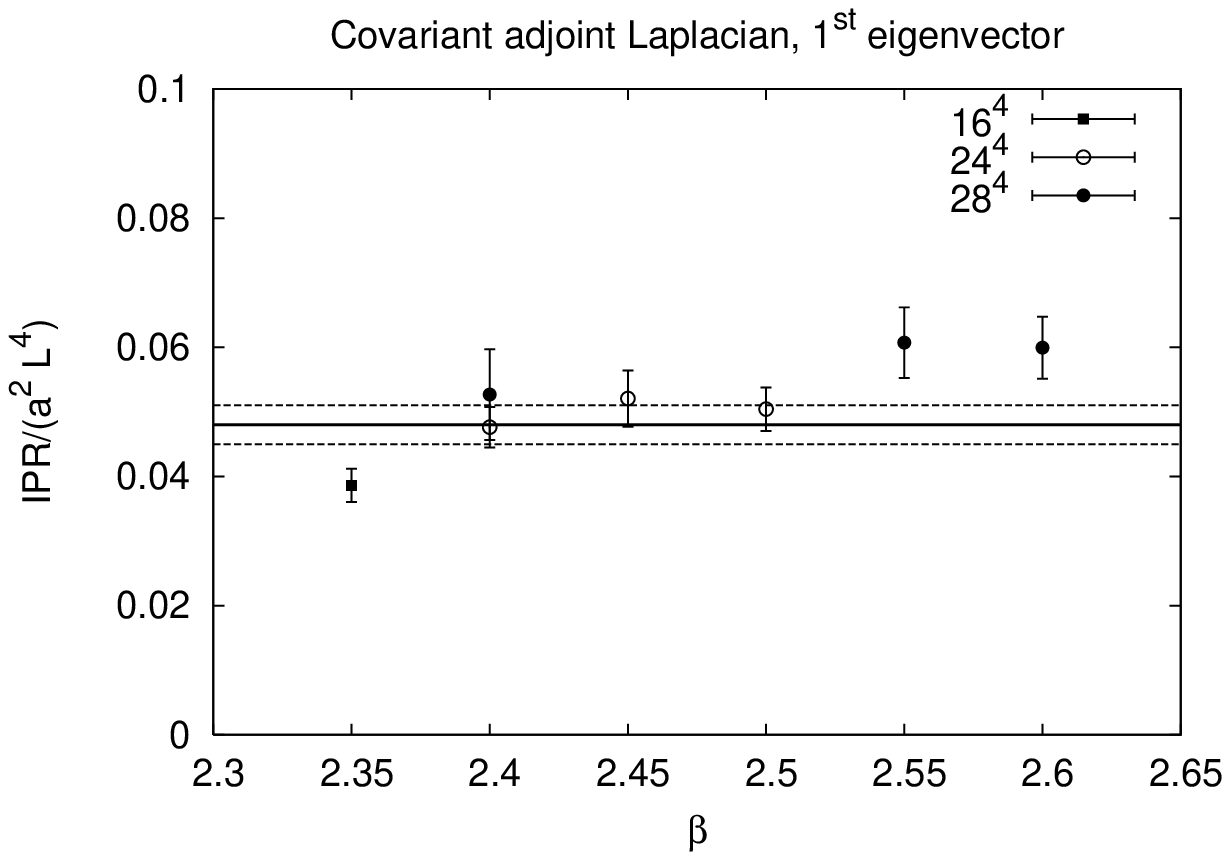}
\ec\end{minipage} \caption{Adjoint Laplacian eigenmodes, zero
temperature.} \label{fig:adjZero}
\end{figure}

    The adjoint covariant transporter is (with $U_{x,\mu}$ being in the
fundamental representation)
\beq \lp[U^{\alpha\beta}_{x,\mu}\rp]_{adj} = {1\over
2}\Tr\lp[U_{x,\mu}\sigma^\alpha
    U_{x,\mu}^{\dag}\sigma^{\beta}\rp]
\eeq
which is $SO(3)$ group-valued and has trivial image of the center
subgroup. IPR values for the lowest eigenmodes (e.m.'s) are shown in
Fig.~\ref{fig:adjZero} which covers a wide range in weak couplings
($\beta = 2.10\ \dots\ 2.60 $) and lattice volumes. The most
striking fact is the scaling of $\IPR$ with lattice spacing $a$:
\beq a^2 \cdot \IPR \sim V_{tot}\,,\qquad V_{loc} \approx const
\cdot a^2\,. \eeq
The shape of the localization region turns out to be approximately
spherical, as clearly seen from density visualizations
\cite{httpLoc}. The radius of support of any localized mode shrinks
to zero as $a \to 0$.

\begin{figure}[t!]
\begin{minipage}{\pghalf}\bc
$V_{loc} \sim a^2 $ \\
\includegraphics[scale=\schalfps]{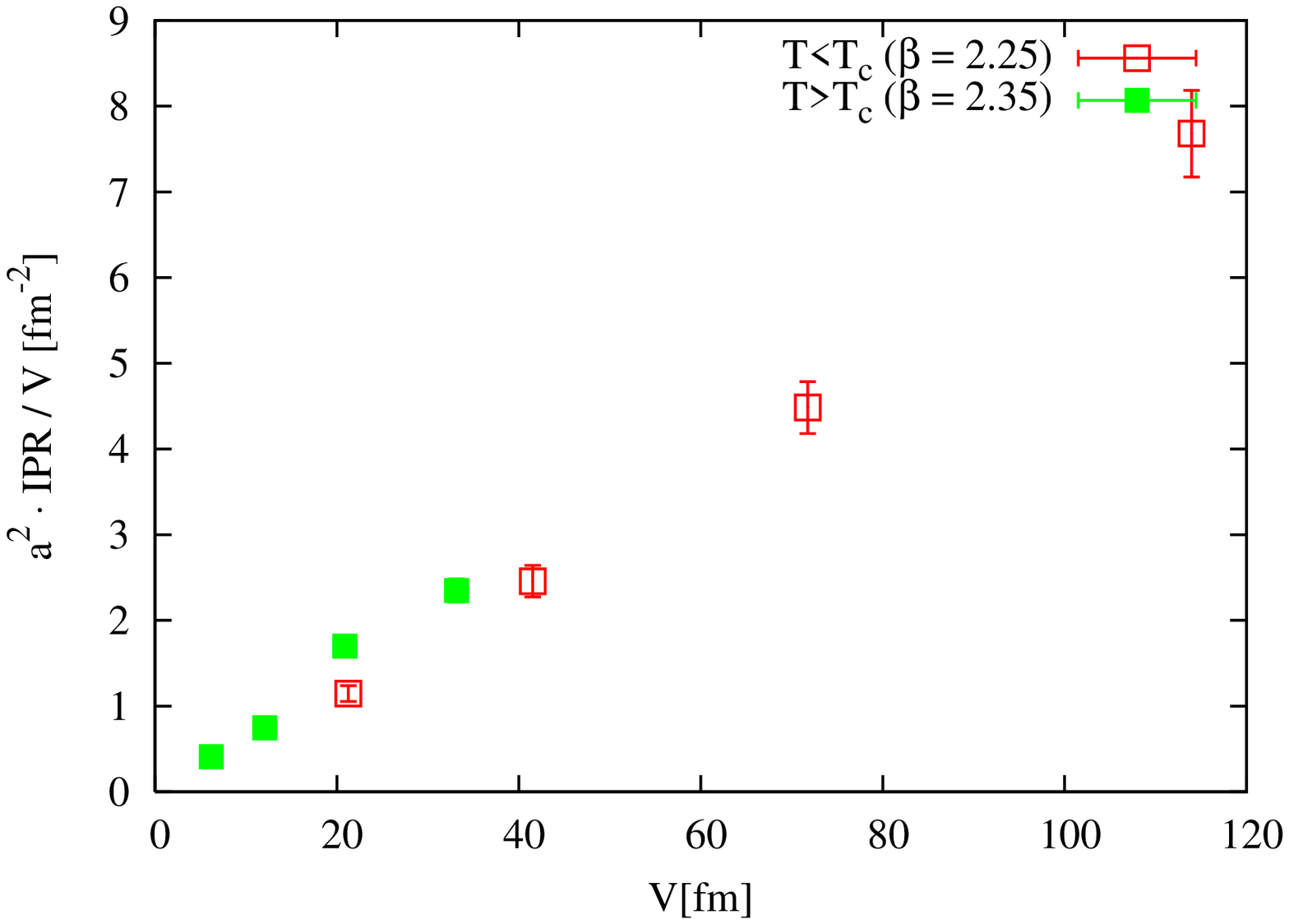}
\ec\end{minipage}
\begin{minipage}{\pghalf}\bc
$V_{loc} / a^2  \approx const $ \\
\includegraphics[scale=\schalfps]{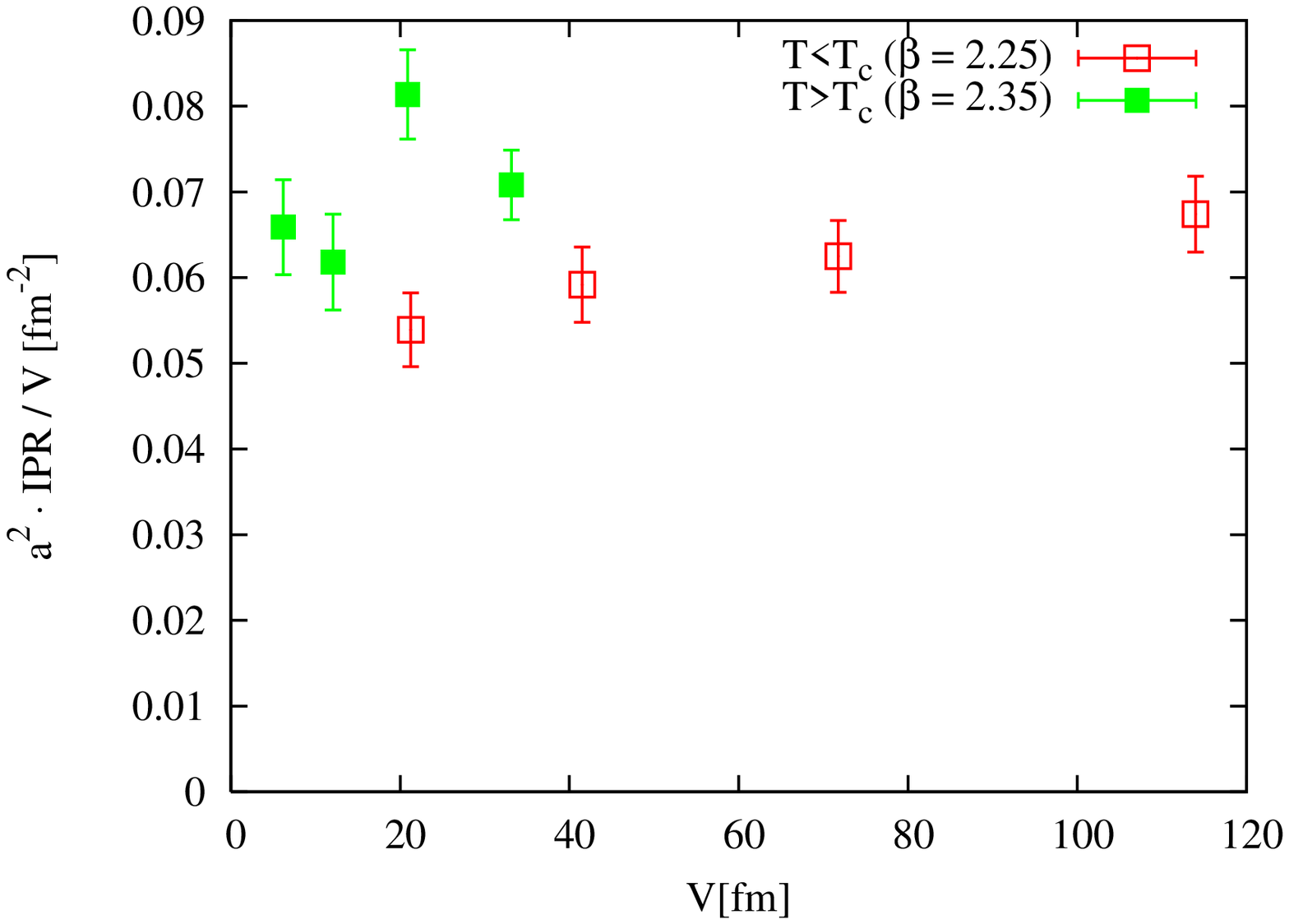}
\ec\end{minipage} \caption{Adjoint Laplacian eigenmodes, below and
above the phase transition.} \label{fig:adjNZ}
\end{figure}

    The same analysis is performed for finite temperature field
configurations. We take lattices with time extension of $L_t = 4$
(the critical point is at $\beta_c \approx 2.30$), while the space
extension of our lattices varies between $L_s = 16\ \dots\ 28$
lattice spacings. To see the effect of crossing the phase transition
we used values of $\beta=2.25$ and $\beta=2.35$. For any point 20
independent configurations are sampled, sufficient to reveal the
qualitative behavior of $\IPR$s. Figures~\ref{fig:adjZero} and
\ref{fig:adjNZ} show the same scaling behavior of $\IPR$, hence the
same scaling of localization volume remains valid both below and
above the deconfinement temperature $T_c$.

    Despite the similarity of results in confinement and deconfinement phases
the localization is related to infrared phenomena. Dimensional analysis of
localization volume implies that
\beq
V_{loc} = const \cdot a^2 = {a^2 \over \LQ^2}
\eeq
and $V_{loc}$ is determined by some mixed scale.

\begin{figure}[t!]
\bc\includegraphics[scale=\schalf]{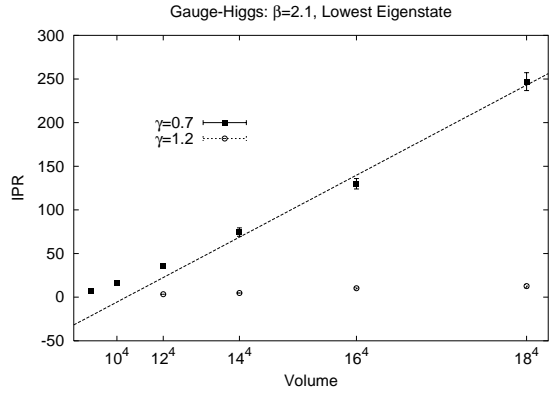}\ec
\caption{Adjoint Laplacian eigenmodes in the gauge-Higgs model, both
confinement-like and Higgs-like phases.} \label{fig:adjHiggs}
\end{figure}

    Now the following question is addressed: Could such localization
result from ordinary gaussian fluctuations, or is it due to
confining features of the quantum vacuum? To check this, we simulate
the model of gauge fields coupled to Higgs fields in the fundamental
representation. It is known to have two phases: confinement-like and
Higgs-like \cite{fradkin,JGcoulomb}, but any two points in the phase
diagram can be joined by a line along which the free energy is
entirely analytic. The transition between the two phases is the
vortex depercolation transition: In the confinement-like phase
vortices are abundant and percolate over the whole lattice volume,
while in the Higgs-like phase the vortex density is small and
vortices do not percolate \cite{Langfeld,Bertle:2003pj}. The model
action is given by
\beq S =
\beta\sum_{plaq}{1\over 2}\Tr\lp[U U U^\dag U^\dag\rp] +
    \gamma\sum_{links}{1\over
    2}\Tr\lp[\phi^\dag(x)U_\mu(x)\phi(x+\hmu)\rp].
\eeq
At $\beta=2.1$ the phase transition occurs at $\gamma=0.9$. Two
values of $\gamma$ are taken for comparison: $\gamma=0.7$
(confinement-like) and $\gamma=1.2$ (Higgs-like).
Fig.~\ref{fig:adjHiggs} shows a drastic reduction of $\IPR$ in the
Higgs phase. Other tests \cite{locpaper} also show that in the
Higgs-like phase the lowest eigenmodes are much more extended; a
crucial point is that the falloff of the density outside the support
is not exponential in the Higgs-like phase, and this is inconsistent with
Anderson localization.

\section{$J=3/2$ representation}

\begin{figure}[b!]
\begin{minipage}{\pghalf}\bc
\includegraphics[scale=\schalf]{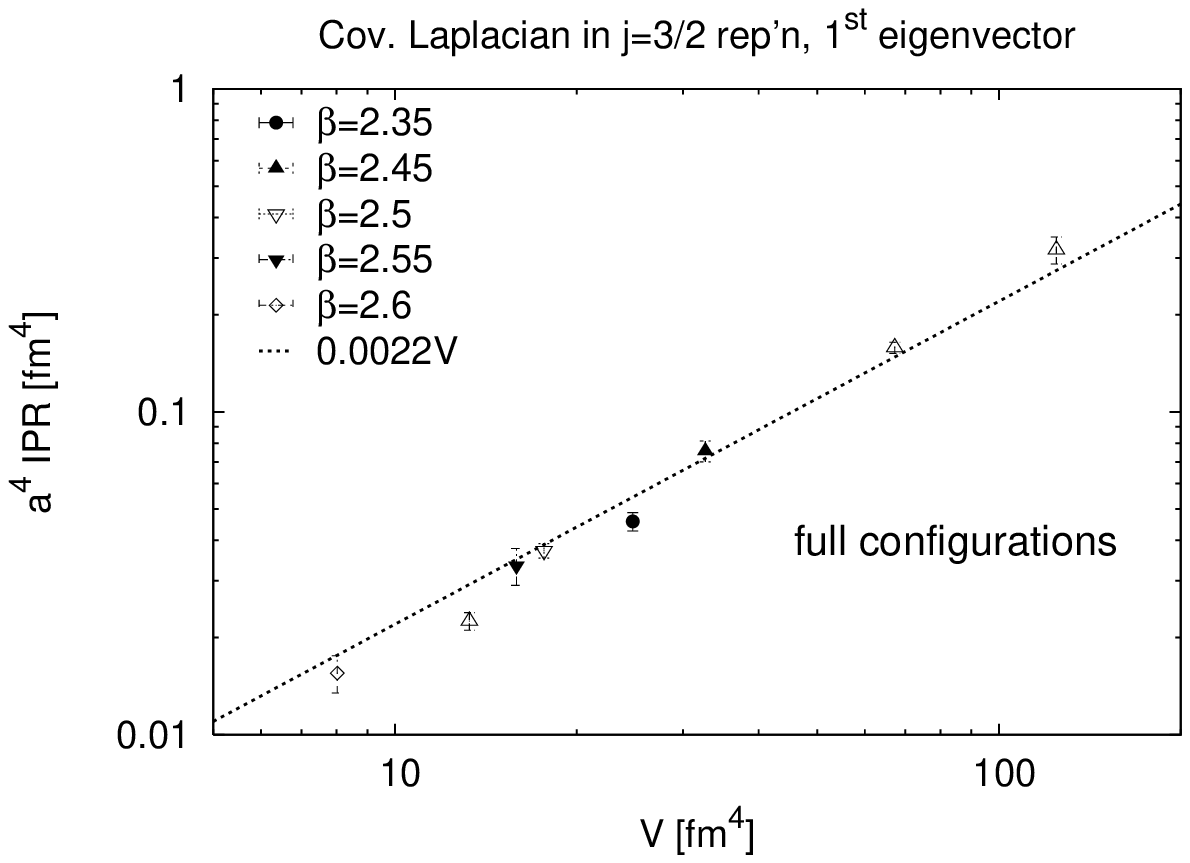} \\
(a) \ec\end{minipage}
\begin{minipage}{\pghalf}\bc
\includegraphics[scale=\schalf]{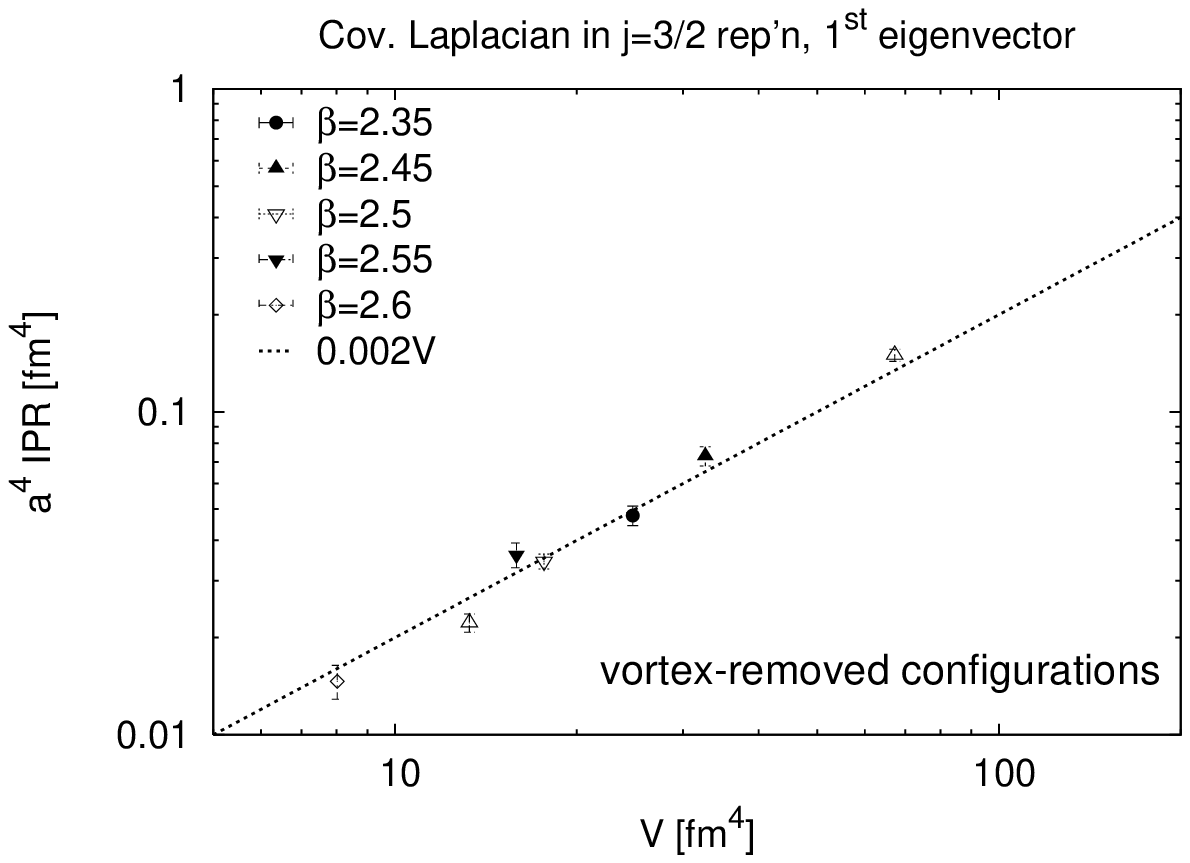} \\
(b) \ec\end{minipage} \caption{Laplacian eigenmodes for $J=3/2$
representation: original gauge (a) and modified (b) fields, zero
temperature.} \label{fig:t3d2Zero}
\end{figure}

    Another representation studied is a complex 4-dimensional, or isospin
$J=3/2$ color SU(2) representation. The center subgroup is
non-trivial and the effect of P-vortices could be separated when one
compares the original gauge field and the one modified via the
de~Forcrand--D'Elia trick~\cite{deForcrand}. The scaling behavior of
$\IPR$ is shown in Fig.~\ref{fig:t3d2Zero}a. The localization volume
diminishes with lattice spacing even more quickly than for the
adjoint representation. From Fig.~\ref{fig:t3d2Zero} one concludes
\beq a^4\cdot \IPR \sim V_{tot}\,,\qquad V_{loc} \approx const\cdot
a^4 \eeq
Density plots in \cite{httpLoc} show that the support of these
localized modes is again spherical. All geometrical parameters of
localization seem to be governed only by the ultraviolet scale
$\LQ$.

    Unlike the case of fundamental representation, the eigenmodes
of $J=3/2$ Laplacian are localized on modified (vortex-removed)
fields almost as sharply as on original fields (see
Fig.~\ref{fig:t3d2Zero}b). This is an evidence in favor of some
other reason for localization than in case of the adjoint and
fundamental covariant Laplacians.

    The same finite-temperature trajectories as for the adjoint representation
were used to study the $J=3/2$ Laplacian eigenmodes. The scaling of
$\IPR$ looks very similar for confinement and deconfinement phases;
it shows that it is ultraviolet fluctuations that are responsible
for localization in this case.

\begin{figure}[t!]
\begin{minipage}{\pghalf}\bc
$V_{loc} \sim a^{\mathbf 4} $ \\
\includegraphics[scale=\schalfps]{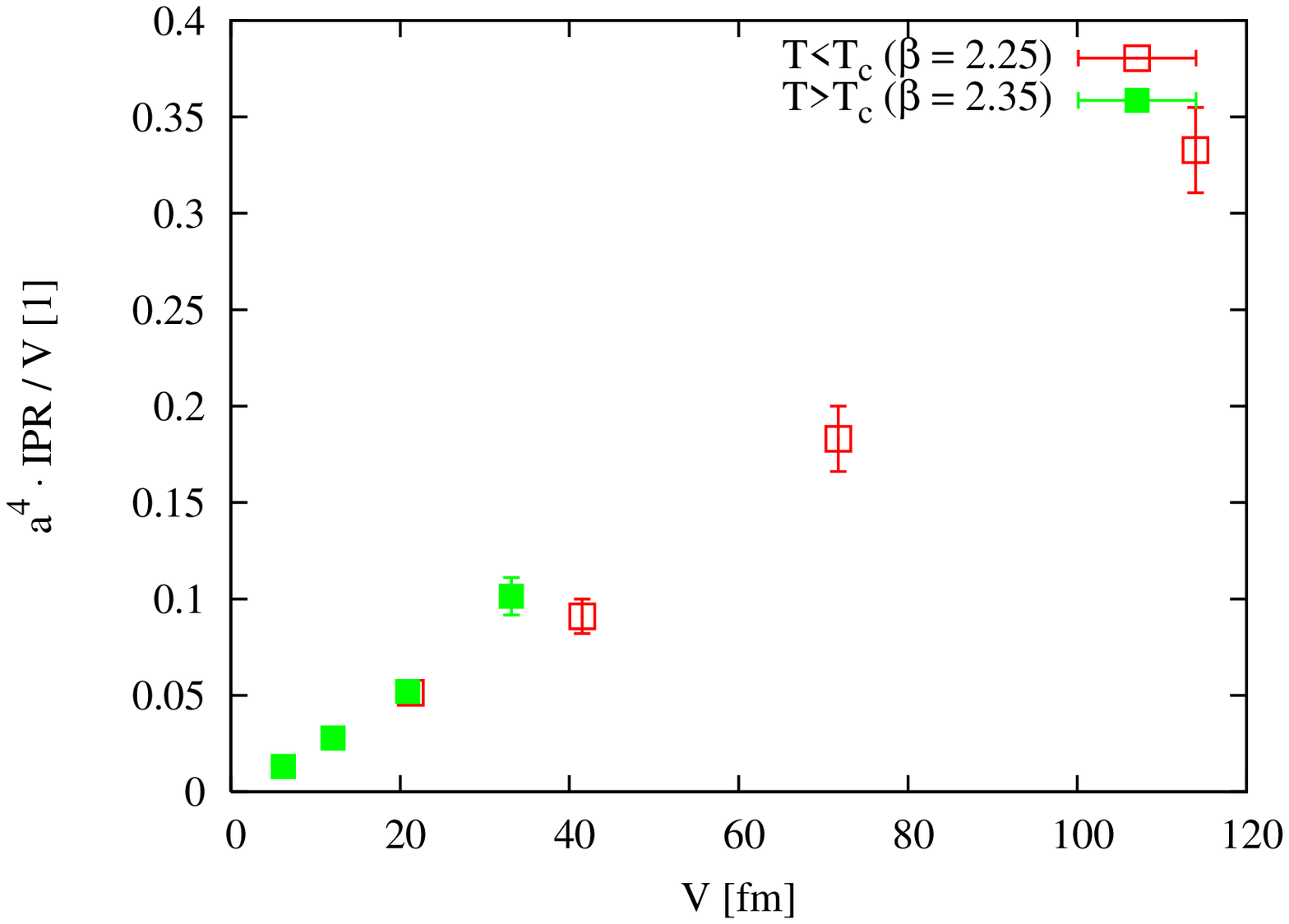}
\ec\end{minipage}
\begin{minipage}{\pghalf}\bc
$V_{loc} / a^4 \approx const$ \\
\includegraphics[scale=\schalfps]{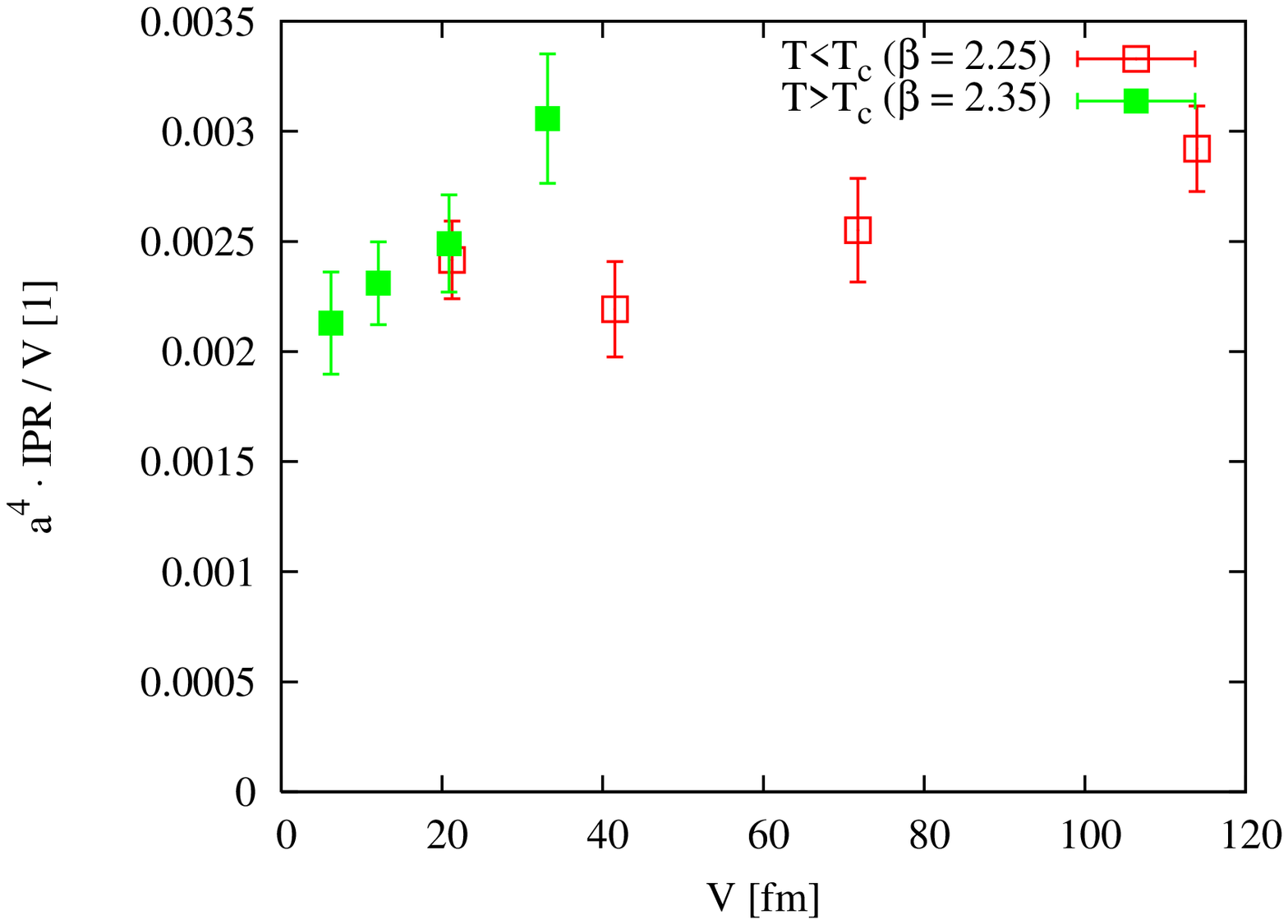}
\ec\end{minipage} \caption{Laplacian eigenmodes for $J=3/2$
representation, finite temperature.} \label{fig:t3d2NZ}
\end{figure}

\section{Summary}

    Naively one would expect similar localization behavior of
eigenmodes of covariant Laplacians in different representations of
the gauge group (up to differences in sizes of localization volumes
due to different interaction strengths), but the naive expectation
is not fulfilled. The presented data demonstrate that at least two
covariant Laplacians in representations other than the fundamental
posess dramatically different features.

    Our results are the following:
\bi
\item The adjoint covariant Laplacian eigenmodes are localized
    in volumes which shrink as $V_{loc} \sim a^2$ in the continuum
    limit. Both scales, infrared and ultraviolet, seem to govern
    the localization.
\item
    Localization in the adjoint representation, as in the
    fundamental representation, is related to the presence of center
    vortices. Vortex removal by the procedure of de Forcrand--D'Elia
    does not work here, but in the gauge-Higgs model, in which the
    vortex content is regulated by the gauge-Higgs coupling
    constant, strong localization is observed in the vortex-abundant
    (confinement-like) phase, while it is absent in the Higgs-like
    phase, where the vortex density is low.
\item The isospin $J=3/2$ representation e.m.'s are localized in
    volumes $V_{loc} \sim a^4$. Vortex removal does not affect this
    phenomenon.
\item The deconfinement transition doesn't influence the character of
    localization in the adjoint and $J=3/2$ representations.
\ei

    The relation of localization of covariant-Laplacian eigenmodes to
confinement is questionable in the light of our results, in
particular of the dependence of the degree of localization on the
group representation of the Laplacian. It may be that localized
eigenmodes in different group representations are probing different
features and different length scales of the QCD vacuum.

%\section*{Parameters of simulation}
%
%\begin{table}
%\begin{tabular}{c|cc|c|c}
%\hline
%$\beta$ &  $L_t$ & $L_s$ & $$a,\,fm$ & $N_{conf}$ \\\hline
%
%\end{tabular}
%\caption{Field trajectories}
%\end{table}

%\section*{Acknowledgements}

\end{document}